\documentclass{mem}
\usepackage{natbib}\usepackage{txfonts}\usepackage{balance}
\usepackage{graphicx}
\usepackage[a4paper]{hyperref}
\idline{76}{1}
\begin{document}
\def\teff{$T\rm_{eff }$}
\def\kms{$\mathrm {km s}^{-1}$}

\title{
The Gaia mission
}
   \subtitle{Pulsating stars with Gaia}
\author{
L. \,Eyer
          }
  \offprints{L.Eyer}

\institute{
Observatoire de Gen\`eve, CH-1290 Sauverny, Switzerland\\
\email{laurent.eyer@obs.unige.ch} }

\authorrunning{Eyer}

\titlerunning{The Gaia mission}

\abstract{
Gaia is a cornerstone mission of the European Space Agency (ESA) selected
in 2000, with a target launch date of 2011. The Gaia mission will perform a
survey of about 1 billion sources brighter than $V=20$.  Its goal is to
provide astrometry leading to parallaxes, proper-motions and positions. Astrometric data are complemented by multicolour photometry and a
spectroscopic instrument (up to $V=17$ mag) to provide astrophysical
parameters (as effective temperatures, surface gravities, metallicities) and radial velocities. The survey will have multi-epoch observations allowing
a detection of the variable sources; objects will be measured as a mean 80/100
times over the 5 year mission length in the two instruments on board,
the Astro-/Spectro-instruments respectively.

The data processing of the mission is a major challenge: as a first step
working groups on different themes were created and have been active. 
One working group has been dedicated to variable stars. A consortium for
the data processing is currently being created, the Data Processing and
Analysis Consortium.

Though several aspects of the Gaia mission are under possible changes, we
review some activities related to the mission in the perspective of
pulsating stars.

}
\maketitle{}

\section{Introduction}

The Gaia survey will reach magnitude $V=20$, and will measure about 1 billion stars of our Galaxy, as it is a survey it will measure as well about 1/2 million QSOs, 1/2 million Solar System objects; extended objects like Galaxies and blended stars will also be measured.
 
The foreseen astrometric precision (that is in the median parallax error) at
the end of the 5 year mission is of 20-25 micro arcsec at mag V=15 for a G2V
star. The errors on position and proper motion have to be multiplied by 0.8 and
0.5 respectively.

In total, there are 21 photometric bands: two full-light magnitudes G and GS,
one Broad Band Photometric system (BBP) composed of 5 bands, one Medium
Band Photometric system (MBP) composed of 14 bands. The G magnitude,
derived from the same signal as the one used for the astrometry, is the
most precise: 1 milli-mag at $V=13-15$ and 0.02 mag at $V=20$.

The Radial Velocity Spectrometer (RVS) will determine the radial velocities.
The wavelength coverage is from 848 to 878 nm with a resolution of 11,500. This wavelength interval contains the Ca II triplet and Paschen lines.
The precision will depend on the star spectral type and apparent magnitude:
it will be about  2 km/s for a G5V star at mag 15. For a full assessment
of the performances of the RVS see \cite{DKetal04}.
\begin{figure*}[t!]
\resizebox{120mm}{!}{\includegraphics[angle=90,clip=true]{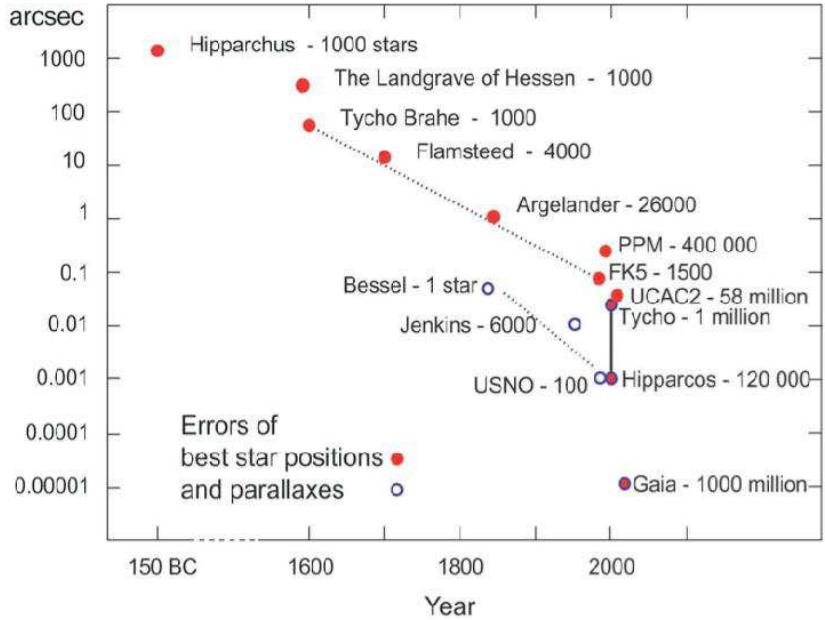}}
\caption{\footnotesize
Astrometric precision throughout history (from E.~H{\o}g).
}
\label{astroprec}
\end{figure*}

If we want to put the Gaia mission in a historical perspective, we remark
that we are witnessing a real revolution. Indeed, if we plot as in
Figure~\ref{astroprec} the astrometric precision as function of centuries,
we note the striking discontinuity which has been started by Hipparcos.
With respect to Hipparcos Gaia is making a jump in precision of nearly 100
for the astrometric precision, and about 10,000 for the number of observed stars.
With Hipparcos, the stars around a sphere in the solar neighbourhood were pinpointed, with Gaia the region explored with high accuracy is a large fraction of our Galaxy.
The progress made during the last 2-3 decades, was made previously during 4 centuries!

The improvement in photometry is also large, from about one full light (Hp) band and the two Tycho bands ($B_T$, $V_T$) we jump to two full light magnitudes,
and 19 bands. In terms of per-transit precision at the bright end, the G-band is about 1 milli-mag with respect to the Hp band which is 4 milli-mag.

At the time of writing this article (2005), two consortia, Alenia/Alcatel
and EADS/Astrium have answered the invitation to tenders by ESA. By the
end of the year 2005 a recommendation will be made and the final design
and construction phase will start by March 2006. We are now in one of the
pivotal eras of the mission. Therefore the given numbers are subject to change. However the numbers presented here have been used as baseline for many studies for several years now.

\section{The instruments on board of Gaia}

The satellite is composed of two independent instruments:\\
~1) The Astro-Instrument: there are two telescopes, two primary mirrors with a rectangular size of 1.4m $\times$ 0.5m. On the focal plane the two telescope viewing directions, the preceding and following fields of view (PFOV, FFOV),
are superimposed. They are separated by a ``basic angle'' of 106 degrees, each covering $0.65 \times 0.65$ deg$^2$.
The focal plane is paved by 170 CCDs, 110 of them are devoted to the Astrometry (AF). A magnitude is extracted from these CCDs, in full light, the so-called G magnitude. The other CCDs are devoted to the Sky Mapper (ASM) and to the Broad Band Photometry (BBP). The CCDs will be operated in Time Delayed Integration mode.\\
~2) The Spectro-Instrument: there is a primary mirror of 0.5m $\times$ 0.5m.
The Spectro focal plane is composed of CCDs devoted to a Sky Mapper (SM), a Radial Velocity Spectrometer (RVS) and a Medium Band Photometry (MBP). The field of view of the spectrograph is $2 \times 1.6$ deg$^2$.

\begin{figure*}[t!]
\resizebox{145mm}{!}{\includegraphics[width=80mm]{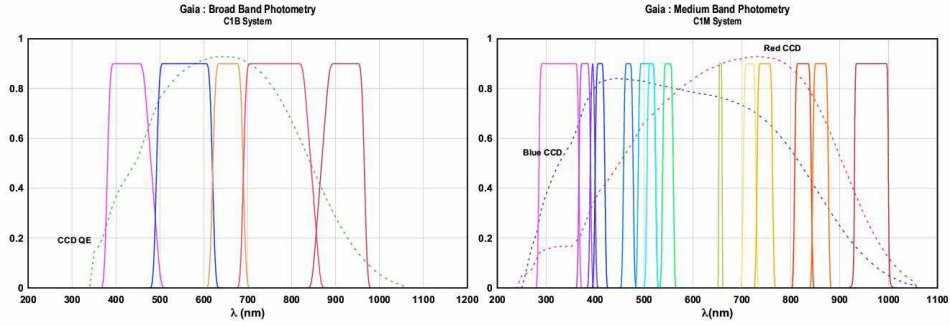}}
\caption{\footnotesize
The two photometric systems, the Broad Band Photometry (BBP) with 5 filters (left), and the Medium Band Photometry (MBP) with 14 filters (right).
}
\label{photometry}
\end{figure*}

In the Figure~\ref{photometry}, we present the bands of the two photometric systems, the BBP (from the Astro focal plane) and the MBP (from the Spectro focal plane).

\section{The time sampling}
The Gaia satellite will evolve around the second Lagrangian point on a Lissajous orbit located at about 1.5 million km from the Earth in direction opposite to the Sun. The choice of the Lissajous orbit will avoid the penumbra of the Earth, which would otherwise cause thermal perturbations. The satellite will rotate on itself in 6 hours. The rotation axis will precess on a cone with an opening angle of 50 degrees to the Sun. This way the satellite will scan the whole celestial sphere. This peculiar scanning law is optimised for astrometric reasons and determines the satellite sampling. The number of measurements will mainly depend on the ecliptic latitude $\beta$, which ranges between 50 to 220 for the Astro-field and for the RVS. The number of measurements of the MBP has changed because the position of the CCDs devoted to the MBP in the Spectro focal plane has changed. This number is now similar to the one obtained in the RVS.

In the Astro-field, the time interval between preceding and following fields of view is 1h46 and between the following and preceding fields of view 4h14. There will be a succession of measurements with such intervals and then larger gaps of about 1 month. 
It will be similar for the Spectro-field though the interval between successive measurements is 6 hours.

\begin{figure}[t!]
\resizebox{70mm}{!}{\includegraphics[angle=90,clip=true]{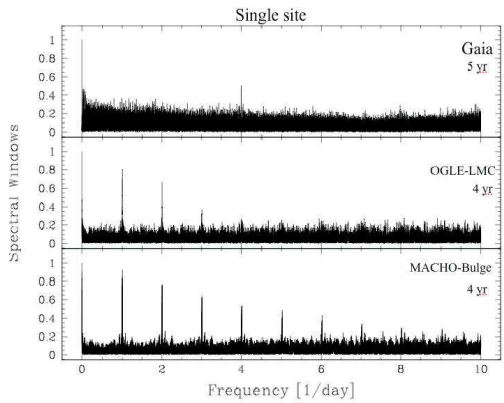}}
\caption{\footnotesize
Spectral windows of Gaia (top), and two other surveys OGLE (middle),
MACHO (bottom). The spectral windows of single sites make them prone to high peaks, which when convolved with the signal will produce aliasing frequencies. The visibility of the bulge makes the spectral window particularly prone to high peaks.
}
\label{spectralwindow}
\end{figure}

For comparison, we plot in Figure~\ref{spectralwindow} three spectral windows of different surveys. The top spectral window is a typical one of Gaia for the Astro-fields of view. We remark that on the frequency interval up to 10 cycles/day, there is the peak of 6 hours which is apparent but not very strong. The global noise level is high because the number of measurements is lower with respect to the two other surveys presented below. Because of the irregular sampling of Gaia there are no high peaks. The middle plot is the spectral window of OGLE for 4 years of observation in the direction of LMC. The well known nightly regular sampling is quite strong. The bottom spectral window is from bulge data of MACHO for 4 years. Because of the bulge sky position, which constrains further the time window during which the bulge is observable, the peaks in the spectral window are higher and will cause stronger aliasing in the power spectrum. The introduction of more randomness in the time sampling when possible may reduce these peaks cf. \cite{LEPB99}.

\section{Gaia and the pulsating stars}

All the aspects of the Gaia mission will contribute to pulsating star knowledge.

\subsection{The astrometry}

Thanks to the measurement of parallaxes, the zero point of the period
luminosity relation of Cepheids,  as well as the absolute magnitudes
of RR Lyrae stars and other standard candles will be determined and
scrutinised to detect any other dependencies.

The absolute magnitude is also crucial for asteroseismologic studies and
will be used. As pointed out by \cite{FF99}, there are two possible approaches
either using the luminosity as a starting point of asteroseismic models or
to use it to confront the predictions obtained from asteroseismologic modelling.

Asteroseismology is mostly done on very bright sources. Therefore the bright
end to which Gaia is obtaining astrometry is important. In the current plan,
this limit is at $G=6$.

\section{The photometry}

Thanks to its multi-epoch photometry and its high precision in the G-band,
Gaia will be able to detect large numbers of pulsating variable stars. 
Some estimations of these numbers have been made \citep{LEJC00}, the harvest will be huge: 3,000 $\beta$ Cephei stars, 15,000 SPBs, 60,000-240,000 $\delta$ Scuti stars, 2,000-8,000 for Cepheids, 70,000 RR Lyrae stars, 200,000 Mira-SR stars.

For periodic signals a study has been made by \cite{LEFM05}. The probability to recover the period of the periodic signal is very high, although it depends on the ecliptic latitude $\beta$. For instance, a sinusoidal signal of 3 milli-mag amplitude at V= 14,
will have its period correctly recovered, this from periods ranging from several hours to hundreds of days.

For irregular variable stars, the situation is more problematic, because of the sparse sampling of the scanning law. However some regions of the sky will be 
regularly sampled over 5 to 10 days.

Very short time scales are also accessible, the photometry is downloaded per CCD
with an integration time of 3.3 seconds. Trends can be detected during the 36 seconds of one star image passage on the 11 CCDs. It is to be studied if EC14026 and roAp stars could be detected. Though Gaia will not be able to describe properly the variability of these sources,  Gaia may be able to detect them.

Thanks to the many bands of Gaia, it will be possible to identify the physical nature of the variability such as pulsation. There also detailed 
studies need to be carried out. Pulsating stars in binary systems (eclipsing binaries) will also detected.

Because of the luminosity derived from the parallax and the colour information,
Gaia will precisely locate the stars in the Hertzprung-Russell (HR) diagram.
The properties of pulsating stars will be studied in detail across the HR
diagram. Gaia will permit to delineate precisely the instability strips, and
the fraction of variable sources in the underlying population for certain
variability thresholds. The effect of chemical composition on pulsation will
be studied for the different variability types.

\section{The radial velocities}
The radial velocity variations for stars with large amplitude can be studied.
For Cepheids and RR Lyrae stars the amplitudes are typically from 10 to 50 km/s, so within the domain of Gaia. With the photometry, radial velocities can be used with the Baade-Wesselink method to estimate masses and radii. For a more extensive discussion on the RVS and variable stars see \cite{MWetal05}.

\section{The data processing for variability}
The size of data downloaded to Earth is estimated to be
about 100-200 Terabytes. The size of the full database on ground is estimated to be
1-2 Petabytes. Because of the interconnectivity of the data, and the different approaches to the data (spatial, temporal, per source, etc\ldots), the data processing is seen as a major challenge. The data processing needs a thorough
preparation. About a dozen working groups were created and now a consortium is being formed.

\subsection{The Variable Star Working Group (VSWG)}
As every Gaia working group, the VSWG is divided into core and associate
members. We count in total about 45 people involved. About half of the
members have been active. The work which has been done covers diverse
subjects, we mention some of them here: a)~We started a uniform description of the characteristics of variable objects (mostly pulsating stars). b)~We also started to introduce variable stars in the Galactic model used for global simulations of the
Gaia data processing. c)~We wrote a code for the detection of variability and tested it using a GRID system.

The future VSWG activities will address several questions:
I)~Period search benchmark. Many period searches are available, their efficiency should be tested with respect to different variability types.
II)~Benchmark of classification methods of astronomical time series. Few works have been published on automated classification methods. However there are several methods available for supervised and unsupervised classifications. We want to estimate their performances.
III)~The usage of Principal Component Analysis on the covariant matrix of the time series in the different bands should be a very efficient way to take advantages of the multi-band photometry of Gaia. It should be studied.
IV)~The variability detected in the Radial Velocity Spectrometer should be studied in detail. \\
For more information please see:
http://obswww.unige.ch/\~\,eyer/VSWG.

\subsection{The Data Processing Analysis Consortium}

Recently the D Analysis Coordination Committee (DACC)
was appointed for structuring a Data Processing and Analysis Consortium. Coordination Units (CU) have been created on different themes: CU1: System Architecture; CU2: Data Simulations; CU3: Core Processing; CU4: Object Processing; CU5: Photometric Processing; CU6: Spectroscopic Processing; CU7: Variability Processing; CU8: Astrophysical Parameters; CU9: Catalogue Access.

\begin{figure*}[t!]
\resizebox{139mm}{!}{\includegraphics[clip=true]{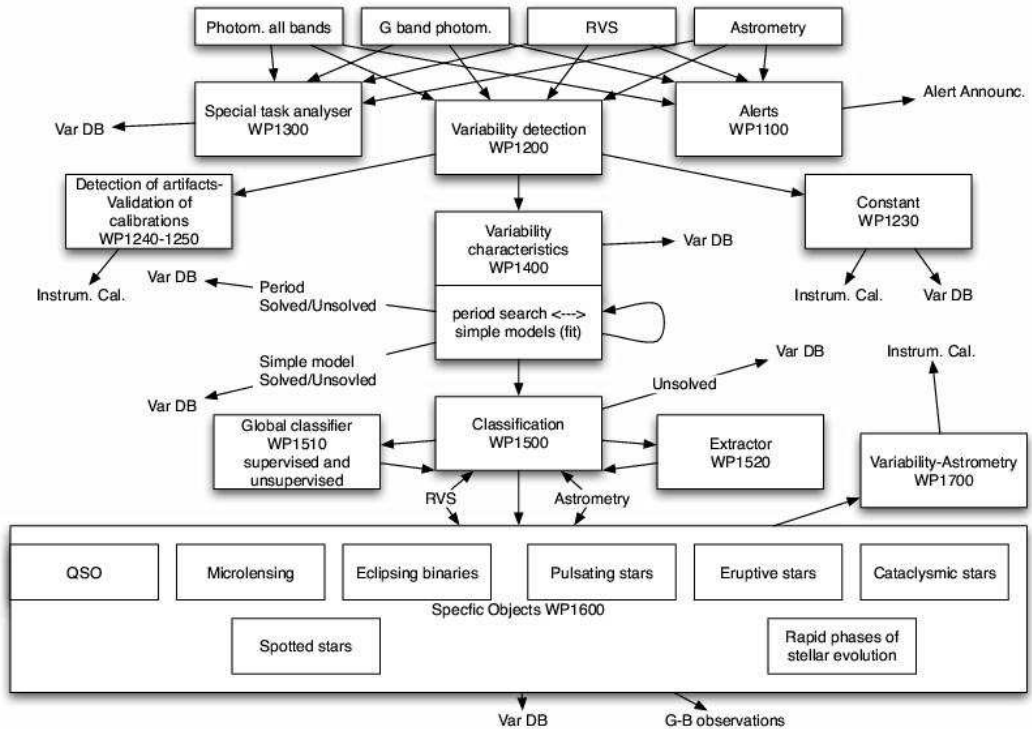}}
\caption{\footnotesize
The different tasks foreseen for the variability analysis. Some are relevant to the photometry
}
\label{va}
\end{figure*}

\subsection{The work breakdown for variability analysis}
In Figure~\ref{va}, we present a possible work breakdown concerning the variability analysis.
The analysis of variability is taking its source in all Gaia information, though
the main source for the detection of variability is the G-band since it is the most precise.  The first step is to detect variability. Then characteristics
of the variability should be defined: as an intrinsic variation amplitude,
a time scale or a period of the variation. Simple models can also be adjusted 
like trend models. Once these characteristics are defined we can populate the
variability database (Var DB in Figure~\ref{va}).  A classification scheme
is to be applied, and variable stars will be classified as pulsating stars.
Variability announcements will be done during the mission.

There is also a box "special task analyser" rooted in all the data whose aim
is the detection of very specific behaviours, which would not otherwise be
flagged as variable by classical tests of variability detection. We just
mention the detection of planetary transits, or the detection of short time pulsators.

All the software has to be tested and optimised. There has been a working group
on Simulation, and there is now a Coordination Unit which is also concerned by
simulations.


\begin{acknowledgements}
I am thankful to Jos de Bruijne for clarifications, Karen O'Flaherty for comments on this text, Dafydd Evans and  the whole Variable Star Working Group.
\end{acknowledgements}
\bibliographystyle{aa}

\end{document}